\documentclass[english,preprint]{aastex}
\usepackage{lmodern}
\usepackage[T1]{fontenc}
\usepackage[latin9]{inputenc}
\usepackage{color}
\usepackage{graphicx}
\usepackage{esint}
\usepackage{babel}

\begin{document}

\title{Do Potential Fields Develop Current Sheets Under Simple Compression
or Expansion?}

\author{Yi-Min Huang\altaffilmark{1,2} and A. Bhattacharjee\altaffilmark{1,2}}

\affil{Space Science Center, University of New Hampshire, Durham, NH 03824}

\author{Ellen G. Zweibel\altaffilmark{2}}

\affil{Department of Physics and Astronomy, University of Wisconsin, Madison,
WI 53706}

\altaffiltext{1}{Center for Integrated Computation and Analysis of Reconnection and Turbulence}

\altaffiltext{2}{Center for Magnetic Self-Organization in Laboratory and Astrophysical Plasmas}

\begin{abstract}
The recent demonstration of current singularity formation by Low \emph{et
al. }assumes that potential fields will remain potential under simple
expansion or compression \citep{Low2006a,Low2007,JanseL2009}. An
explicit counterexample to their key assumption is constructed. Our
findings suggest that their results may need to be reconsidered.
\end{abstract}

\section{Introduction}

The theory of force-free magnetic fields is of great interest in many
astrophysical situations \citep{Marsh1996,Sturrock1994}. For instance,
in the quiescent solar corona, the magnetic energy dominates the thermal
energy, the kinetic energy, and the gravitational potential energy.
Under these conditions, the Lorentz force $\mathbf{J}\times\mathbf{B}$
of the magnetic field is approximately zero, therefore the field is
force-free to the lowest order. A force-free field satisfies \begin{equation}
\mathbf{J}=\nabla\times\mathbf{B}=\lambda\mathbf{B};\label{eq:FFa}\end{equation}
that is, the electric current \textbf{$\mathbf{J}$ }is parallel to
the magnetic field $\mathbf{B}$ everywhere. From the condition$\nabla\cdot\mathbf{B}=0$,
we obtain $\mathbf{B}\cdot\nabla\lambda=0.$

The magnetic field lines in the solar coronal loops are deeply anchored
in the dense photosphere, which is continually moving due to the convective
motions in the solar interior. As a result, the coronal magnetic field
is constantly changing in response to the photospheric motions. Since
the typical turnover time of photospheric motions ($\sim$hundreds
of seconds) is much longer than the Alfv\'en transit time ($\sim$tens
of seconds), the solar coronal loops remain approximately force-free
for all time, as long as they are quiescent. In other words, to a
good approximation, the photospheric motions carry the coronal magnetic
field through a sequence of quasi-static, force-free equilibria. It
is, of course, possible that the force-free field may become unstable
to magnetohydrodynamic (MHD) instabilities. When instabilities occur,
the coronal loops will no longer remain quasi-static, and subsequently
some violent events, such as coronal mass ejections (CMEs), may ensue. 

The solar corona is highly conducting. To a very good approximation,
the plasma resistivity is negligible. One of the most important problems
in solar physics is to explain how the solar corona could be heated
to millions of degrees when the resistive dissipation seems so weak
\citep{Klimchuk2006}. Parker proposed a scenario, which he termed
{}``topological dissipation'', as a possible mechanism of coronal
heating \citep{Parker1972,Parker1983,Parker1979,Parker1994}. According
to Parker, the magnetic field in an ideal plasma has a natural tendency
to form tangential discontinuities, that is, current singularities,
when the field line topology is rendered sufficiently complicated
by the boundary motions of the photosphere. In reality, because of
the finite resistivity, the ideal current singularities will be smoothed
out and become thin current filaments. The current filaments are so
intense that even the tiny resistivity of solar corona can lead to
significant dissipation. Parker suggests that the dissipation from
ubiquitous thin current filaments could be the energy source of coronal
heating.

Parker's claim has stimulated considerable debate that continues to
this day \citep{vanBallegooijen1985,ZweibelL1987,Antiochos1987,LongcopeS1994a,NgB1998,CraigS2005},
without apparent consensus. Recently, in a series of papers \citep{Low2006a,Low2007,JanseL2009},
Low and coauthors try to demonstrate unambiguously the formation of
current singularities using what they call {}``topologically untwisted
fields'', which, in the context of force-free field, is synonymous
with potential fields. That is, they limit themselves to a special
subset of force-free fields with $\lambda=0$ in Eq. (\ref{eq:FFa}),
which correspond to current-free fields. In this case, the magnetic
field can be expressed as $\mathbf{B}=\nabla\chi$ with some potential
$\chi$, where $\chi$ satisfies $\nabla^{2}\chi=0$ because $\mathbf{B}$
is divergenceless. In a simply connected, compact domain, $\chi$
is uniquely determined up to an additive constant if the normal derivative
of $\chi$ is given on the boundary. Therefore, a potential field
is uniquely determined by prescribing the normal component of $\mathbf{B}$
($=\hat{\mathbf{n}}\cdot\nabla\chi$, where $\hat{\mathbf{n}}$ is
the unit normal vector) on the boundary. \citet{JanseL2009} consider
a potential field in a cylinder of finite length as an initial condition;
the normal component of the field is nonzero only on the top and the
bottom of the cylinder. The cylinder is then compressed to a shorter
length. Because the whole process is governed by the ideal induction
equation, the normal component of the field remains the same. By making
a \emph{key} \emph{assumption} that the field will remain potential
during the process, they circumvent the complicated problem of solving
for the various stages of quasi-static evolution, and simply calculate
the magnetic field in the final state as the potential field which
satisfies the boundary conditions. They then numerically calculate
the magnetic field line mapping from one end to the other, for both
the initial and the final states. They find that for a three-dimensional
(3D) field, the field line mapping, and hence, the topology, is changed.
They claim that current singularities must form during the process;
furthermore, singularities may form densely due to the ubiquitous
change of the field line mapping. 

The assumption that the field will remain potential during the compression,
however, is not proven in their work. Intuitively it looks plausible,
as parallel current is related to the relative twist between neighboring
field lines. It would appear to require a nonzero vorticity on the
boundary, which is absent in a simple expansion or compression, to
twist up the field. However, parallel current is related to the relative
twist between neighboring field lines only in an average sense; the
relative twist between a specific pair of field lines is actually
independent of the parallel current. Although evolutions following
the ideal induction equation preserve the footpoint mapping of each
individual field line, it does not follow that parallel current will
not be generated during a simple expansion or compression. In this
paper an explicit counterexample is constructed in a two-dimensional
(2D) slab geometry; therefore the demonstration of current singularity
formation by Low \emph{et al. }is put in doubt.

\section{Two-dimensional Counterexample}

Consider a 2D configuration in Cartesian geometry, with $x$ the direction
of symmetry. A general magnetic field may be expressed as \begin{equation}
\mathbf{B}=B_{x}\mathbf{\hat{x}}+\mathbf{\hat{x}}\times\nabla\psi,\label{eq:2d_B}\end{equation}
where $B_{x}=B_{x}(y,z)$ and $\psi=\psi(y,z)$. For a force-free
field, $B_{x}=B_{x}(\psi)$ and $\psi$ must satisfy the Grad-Shafranov
equation:\begin{equation}
\nabla^{2}\psi=-B_{x}\frac{dB_{x}}{d\psi}.\label{eq:GS}\end{equation}
The $\lambda$ for a force-free field in Eq. (\ref{eq:FFa}) is equal
to $-dB_{x}/d\psi$ in this representation. If the field is potential,
then $B_{x}=\mbox{const}$ and $\nabla^{2}\psi=0$. 

Let us consider a force-free field enclosed by two conducting boundaries
at $z=0$ and $z=L$, with all field lines connecting one end plate
to the other. The footpoint mapping from one end to the other is characterized
by $\psi$ on the boundaries, as well as the axial displacement along
$x$ when following a field line from the bottom to the top: \begin{equation}
h(\psi)=B_{x}(\psi)\left(\int_{z=0}^{z=L}\frac{1}{\left|\nabla\psi\right|}ds\right)_{\psi}.\label{eq:h}\end{equation}
Here $ds$ is the line element on the $y-z$ plane, and the subscript
$\psi$ indicates that the integration is done along a constant $\psi$
contour. Notice that $h(\psi)=d\Phi/d\psi$, with $\Phi(\psi)$ the
axial magnetic flux function up to an additive constant. The displacement
function $h(\psi)$ resembles the safety factor $q(\psi)$ in toroidal
geometry. 

Let the initial condition be $(B_{x0},\psi_{0})$ with a system length
$L_{0}$. Suppose the system has undergone a simple expansion or compression
such that the system length changes to some $L\neq L_{0}$. Since
ideal evolution preserves the footpoint mapping, the flux function
$\psi$ on the boundaries and the axial displacement $h(\psi)$ must
remain unchanged. To determine the final force-free equilibrium, we
have to solve the two coupled equations \begin{equation}
\nabla^{2}\psi=-B_{x}\frac{dB_{x}}{d\psi}\label{eq:GS1}\end{equation}
\begin{equation}
B_{x}(\psi)=h(\psi)\left(\int_{z=0}^{z=L}\frac{1}{\left|\nabla\psi\right|}ds\right)_{\psi}^{-1}\label{eq:Bx2}\end{equation}
subject to $\psi(y,0)=\psi_{0}(y,0)$, $\psi(y,L)=\psi_{0}(y,L_{0})$,
where the axial displacement $h(\psi)$ is determine from the initial
condition. The set of coupled equations (\ref{eq:GS1}) and (\ref{eq:Bx2})
is called a generalized differential equation, which in general requires
numerical solutions \citep{GradHS1975}. 

As a simple example, consider the following initial condition: $\psi_{0}=y+\epsilon\sin(y)\cosh(z-1/2),$
$B_{x0}=1$, and $L_{0}=1$, where $\epsilon\ll1$ is a small parameter.
This initial field is potential. Now the question is, will the final
field remain potential when the system is expanded or compressed to
another length $L$? We may first assume that it will, and see if
that leads to a contradiction. Assuming that, we have to solve $\nabla^{2}\psi=0$
subject to boundary conditions $\psi(y,0)=\psi(y,L)=y+\epsilon\sin(y)\cosh(1/2)$,
and $B_{x}=\mbox{const}$. The solution for $\psi$ is \begin{equation}
\psi=y+\epsilon a\sin(y)\cosh(z-L/2),\label{eq:sol_pot}\end{equation}
where \begin{equation}
a=\frac{\cosh(1/2)}{\cosh(L/2)}.\label{eq:a}\end{equation}
The constant $B_{x}$ may be determined by the conservation
of axial flux, which yields $B_{x}=1/L$. To be consistent with the
initial footpoint mapping, the final state $(\psi,B_{x})$ must give
the same axial displacement as $(\psi_{0},B_{x0})$. We examine this
along the field line $y=0$, corresponding to $\psi_{0}=\psi=0$.
For the initial field,\begin{eqnarray}
h(\psi_{0}=0) & = & \int_{0}^{1}\frac{1}{1+\epsilon\cosh(z-1/2)}dz\nonumber \\
 & = & \int_{-1/2}^{1/2}\frac{1}{1+\epsilon\cosh z'}dz';\label{eq:h1}\end{eqnarray}
and for the final field, \begin{eqnarray}
h(\psi=0) & = & \frac{1}{L}\int_{0}^{L}\frac{1}{1+\epsilon a\cosh(z-L/2)}dz\nonumber \\
 & = & \int_{-1/2}^{1/2}\frac{1}{1+\epsilon a\cosh(Lz')}dz'.\label{eq:h2}\end{eqnarray}
The condition that the axial displacement remains unchanged implies
\begin{equation}
\int_{-1/2}^{1/2}\frac{a\cosh(Lz')-\cosh z'}{\left(1+\epsilon\cosh z'\right)\left(1+\epsilon a\cosh(Lz')\right)}dz'=0.\label{eq:condition}\end{equation}
However, it may be shown that the integrand of Eq. (\ref{eq:condition})
for $z'\in(-1/2,1/2)$ is positive when $L<1$ and negative when $L>1$;
therefore the condition (\ref{eq:condition}) cannot be satisfied
except for the trivial case $L=L_{0}=1$ and we have a contradiction.

Having shown that a potential field cannot be the final force-free
equilibrium, the next question is whether a smooth, non-potential
equilibrium exists. To demonstrate the existence of a smooth solution,
let us consider a perturbative solution \begin{equation}
\psi=y+\epsilon\psi_{1}+O(\epsilon^{2}),\label{eq:psi}\end{equation}
with the following \emph{ansatz\begin{equation}
\psi_{1}=\sin(y)g(z).\label{eq:psi1}\end{equation}
}From the initial condition, we have\begin{equation}
h(\psi)=1-2\epsilon\sinh\frac{1}{2}\cos(\psi)+O(\epsilon^{2}).\label{eq:h5}\end{equation}
And Eq. (\ref{eq:Bx2}), to $O(\epsilon)$, yields

\begin{eqnarray}
B_{x}(\psi) & = & h(\psi)\left(\int_{z=0}^{z=L}\frac{1}{\left|\nabla\psi\right|}ds\right)_{\psi}^{-1}\nonumber \\
 & = & h(\psi)\left(L-\epsilon\cos\psi\int_{0}^{L}g(z)dz+O(\epsilon^{2})\right)^{-1}\nonumber \\
 & = & \frac{1}{L}+\frac{\epsilon\cos\psi}{L}\left(\frac{1}{L}\int_{0}^{L}g(z)dz-2\sinh\frac{1}{2}\right)+O(\epsilon^{2}).\label{eq:Bx_b}\end{eqnarray}
The right hand side (RHS) of Eq. (\ref{eq:GS1}) is \begin{eqnarray}
-B_{x}\frac{dB_{x}}{d\psi} & = & \frac{\epsilon}{L^{2}}\sin\psi\left(\frac{1}{L}\int_{0}^{L}g(z)dz-2\sinh\frac{1}{2}\right)+O(\epsilon^{2})\nonumber \\
 & = & \frac{\epsilon}{L^{2}}\sin y\left(\frac{1}{L}\int_{0}^{L}g(z)dz-2\sinh\frac{1}{2}\right)+O(\epsilon^{2}).\label{eq:RHS}\end{eqnarray}
Hence, Eq. (\ref{eq:GS1}), to $O(\epsilon)$, is\begin{equation}
g''(z)-g(z)=\frac{1}{L^{2}}\left(\frac{1}{L}\int_{0}^{L}g(z)dz-2\sinh\frac{1}{2}\right).\label{eq:geq}\end{equation}
And the boundary condition for $\psi$ requires $g(0)=g(L)=\cosh(1/2)$.
The solution can be readily found as \begin{equation}
g(z)=c_{1}\cosh\left(z-\frac{L}{2}\right)+c_{2},\label{eq:g}\end{equation}
where \begin{equation}
c_{1}=\frac{L\left(L^{2}+1\right)\cosh(1/2)-2L\sinh(1/2)}{L\left(L^{2}+1\right)\cosh(L/2)-2\sinh(L/2)},\label{eq:c1}\end{equation}
and\begin{equation}
c_{2}=\cosh\frac{1}{2}-c_{1}\cosh\frac{L}{2}.\label{eq:c2}\end{equation}
Therefore, we have shown that a smooth solution could indeed be found,
at least to $O(\epsilon)$. From Eq. (\ref{eq:Bx_b}), the axial magnetic
field is \begin{equation}
B_{x}(\psi)=\frac{1}{L}+\frac{\epsilon\cos\psi}{L}\left(\frac{2}{L}c_{1}\sinh\frac{L}{2}+c_{2}-2\sinh\frac{1}{2}\right)+O(\epsilon^{2}).\label{eq:Bx_final}\end{equation}
And finally, \begin{equation}
\lambda=-\frac{dB_{x}}{d\psi}=\frac{\epsilon\sin\psi}{L}\left(\frac{2}{L}c_{1}\sinh\frac{L}{2}+c_{2}-2\sinh\frac{1}{2}\right)+O(\epsilon^{2}).\label{eq:lambda_final}\end{equation}
Since $\lambda\neq0$ almost everywhere when $L\neq1$, a smooth electric
current is generated through out the whole domain. 

\begin{figure}
\begin{centering}
\includegraphics[scale=0.85]{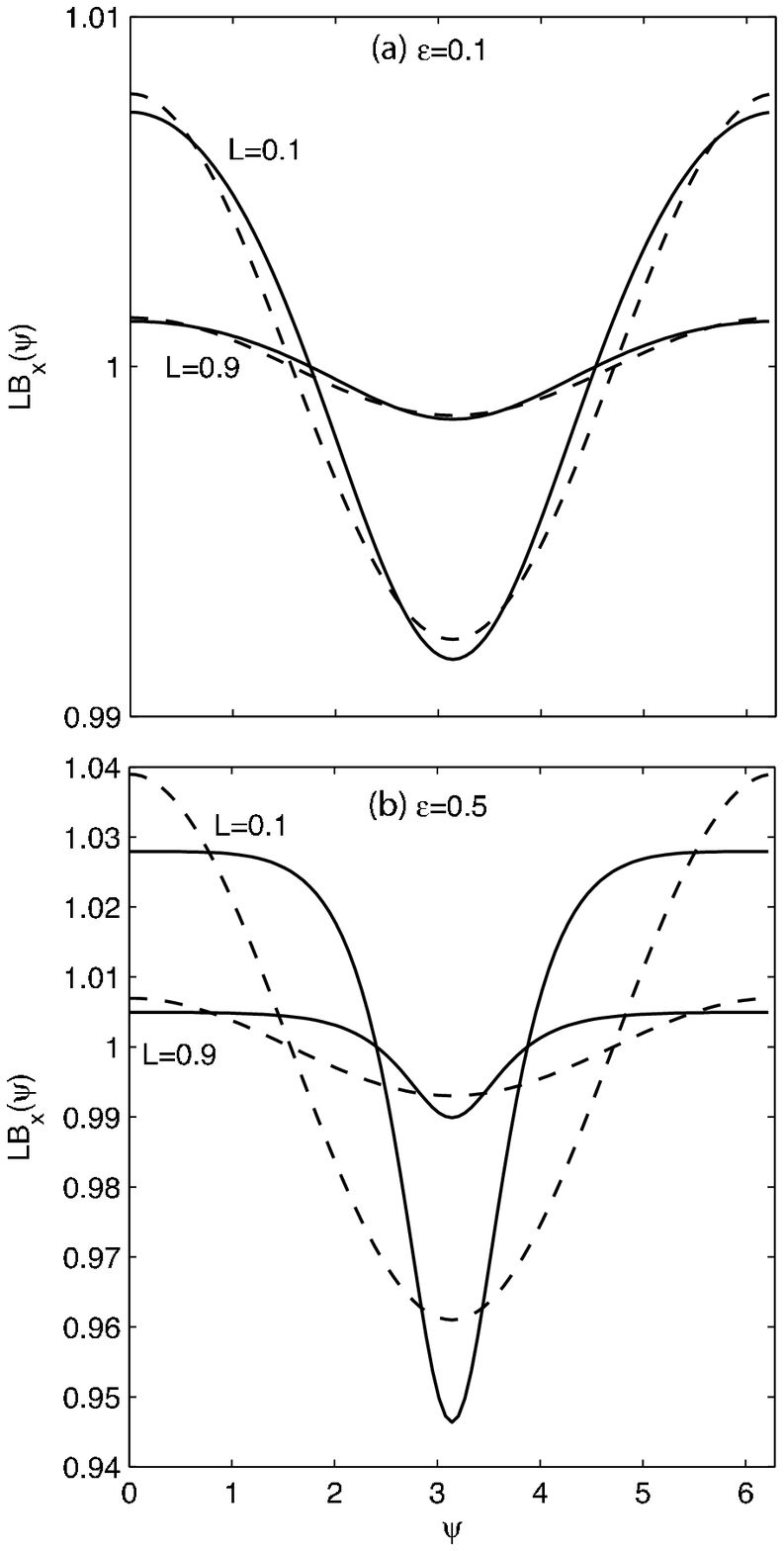}
\par\end{centering}

\caption{$LB_{x}(\psi)$ as a function of $\psi$ for two representative length
$L=0.1$ and $L=0.9$. Panel (a) shows the case $\epsilon=0.1$ and
panel (b) shows the case $\epsilon=0.5$. Solid lines are numerical
solutions and dashed line are perturbative solutions. The $\epsilon=0.1$
case shows good agreement between the perturbative and numerical solutions.
For $\epsilon=0.5$ the agreement is only qualitative. Notice that
$B_{x}=1$ when $L=1$. Since $B_{x}(\psi)$ becomes non-constant
for $L\neq1$, the field becomes non-potential. \label{fig1}}

\end{figure}

We have checked the perturbative solution by the numerical solution
of the generalized differential equation. We find good agreement for
small $\epsilon$. For larger $\epsilon$ the numerical solution does
not quite agree with the solution, as expected. Furthermore, we have
been able to find smooth numerical solutions for a wide range of $L$
and $\epsilon$. Figure \ref{fig1} shows $LB_{x}(\psi)$ as a function
of $\psi$ for two representative lengths $L=0.1$ and $L=0.9$, with
$\epsilon=0.1$ and $\epsilon=0.5$. Solid lines are numerical solutions
and dashed line are perturbative solutions. The $\epsilon=0.1$ case
shows good agreement between the perturbative and numerical solutions.
For $\epsilon=0.5$ the agreement is only qualitative.

\section{Summary and Discussion}

In summary, we have constructed an explicit counterexample to the
assumption that potential fields remain potential under simple expansion
or compression. For our 2D system, the reason that in general the
field will not remain potential is quite simple. A potential field
has to satisfy $\nabla^{2}\psi=0$, which uniquely determines $\psi$
for the given boundary value of $\psi$. This $\psi$, however, has
to satisfy an extra constraint to be self-consistent. That is, when
substituting $\psi$ into the RHS of (\ref{eq:Bx2}), we should get
a constant. This is a very stringent constraint and in general will
not be satisfied, as we have demonstrated with a simple example. 

It should be pointed out that the domain under consideration is not
compact, therefore specifying the normal component of the magnetic
field on the boundary does not uniquely determine the potential field.
One can, for example, add a constant $B_{y}$ or $B_{x}$ to the field
without changing the normal component of $\mathbf{B}$. In this regard,
our configuration does not belong to the same class of configurations
as considered by Low \emph{et al}. However, the purpose of this work
is to demonstrate that a potential field can evolve into a non-potential
field by simply changing the system length, therefore it should not
be taken for granted that the final field has to remain potential.
Since this is the key assumption in the demonstration of current singularity
formation by Low \emph{et al.}, our findings suggest that their conclusion
needs to be revisited. Furthermore, Eq. (\ref{eq:lambda_final}) shows
that a parallel current is generated almost everywhere. This may account
for why Low and coauthors find ubiquitous change in the footpoint
mapping. Although our analytic demonstration of the existence of a
smooth solution is based on perturbation theory, the demonstration
is supported by numerical solution even when perturbation theory is
not strictly valid.

The question posed by title {}``do potential fields develop current
sheets under simple compression or expansion?'' remains open at this
point. Indeed, we cannot preclude the possibility that current sheets
may form during the process. However, even if that is the case, a
smooth current density is likely to precede the formation of current
sheets. This raises the overall question as whether limiting to potential
fields is a viable option in solving Parker's problem. The answer
is probably no.

\acknowledgements{Yi-Min Huang is grateful to Dr. B. C. Low for his hospitality when
YMH visited the High Altitude Observatory and for discussions of his
own work, which inspired this paper. We thank the anonymous referee
for an insightful suggestion on the role of axial flux conservation.
This research is supported by the Department of Energy, Grant No.
DE-FG02-07ER46372, under the auspice of the Center for Integrated
Computation and Analysis of Reconnection and Turbulence (CICART) and
the National Science Foundation, Grant No. PHY-0215581 (PFC: Center
for Magnetic Self-Organization in Laboratory and Astrophysical Plasmas). }

\newpage{}

\bibliographystyle{/Users/yop/Library/texmf/tex/latex/apj}

\end{document}